# Improper flexoelectricity in hexagonal rare-earth ferrites

Xin Li[1†], Guodong Ren[2†], Yu Yun[1], Arashdeep Singh Thind[2], Amit Kumar Shah[1], Abbey Bowers[1], Rohan Mishra[3,2*], Xiaoshan Xu[1,4*]

[1]Department of Physics and Astronomy, University of Nebraska, Lincoln, Nebraska 68588, USA
[2]Institute of Materials Science & Engineering, Washington University in St. Louis, St. Louis MO, USA
[3]Department of Mechanical Engineering & Materials Science, Washington University in St. Louis, St. Louis MO, USA
[4]Nebraska Center for Materials and Nanoscience, University of Nebraska, Lincoln, Nebraska 68588, USA

†These authors contribute equally

*Corresponding author: X.X. (xiaoshan.xu@unl.edu), R.M. (rmishra@wustl.edu)

**Abstract:**

Flexoelectricity is a universal effect that generates electric polarization due to broken inversion symmetry caused by local strain gradient. The large strain gradient at nanoscale makes flexoelectric effects, especially in nanoscopic ferroelectric materials, promising in sensors, actuator, energy harvesting, and memory applications. In this work, we studied flexoelectricity in hexagonal ferrites h-YbFeO$_3$, an improper ferroelectric expected to have weak piezoelectricity and low sensitivity to depolarization field, which are advantageous for studying intrinsic flexoelectric effects. We show that in h-YbFeO$_3$ epitaxial thin films, strain gradient is on the order of $10^6$ m$^{-1}$ occurs near grain boundaries and edge dislocation, which has a significant impact on the non-polar K$_3$ structural distortion that induces spontaneous polarization. The phenomenological model based on the Landau theory of improper ferroelectricity suggests an indirect flexoelectric effect on the order of 10 nC/m in h-YbFeO$_3$, which is substantially larger than the expectation from Kogan's mechanism. These results reveal a novel microscopic mechanism of coupling between strain gradient and polarization mediated by structural distortion, which we call improper flexoelectricity.

Flexoelectricity is a fundamental electromechanical effect characterized by the breaking of inversion symmetry due to a strain gradient, leading to the generation of electric polarization.[1–7] It is universal to any dielectric material, unlike piezoelectricity, which is the coupling between uniform strain and polarization, and is only displayed by non-centrosymmetric crystals. Bulk materials normally have negligibly small strain gradients due to their large strain relaxation length.[8,9] Therefore, their flexoelectric response is negligible.[10–17] On the nanoscale, however, sizable flexoelectric effects are observed due to the large strain gradients imposed by nonuniform distortions.[18-30] In particular, strong flexoelectric effects in ferroelectric materials, which exhibit spontaneous polarization, enable mechanical switching of local polarization.[24–26,31,32] These promising effects have spurred a number of proposed applications, including as electromechanical sensors, actuators, nanogenerators[30], and devices for information storage.[2,3]

Most studies on flexoelectricity in ferroelectric materials have been focused on perovskite materials showing proper ferroelectricity.[8,18,19,21–23,31–34] In this context, the flexoelectric coefficients, defined as $\mu=P/\eta$, are often estimated using Kogan's relation in terms of an electric potential difference caused by a strain gradient $\eta$ and the subsequently induced polarization $P$.[35] This relation can not only estimate the flexoelectric response of bulk ferroelectrics, it can also explain the large $\mu$ observed in relaxor ferroelectrics, wherein the nanopolar domains give rise to a large $\eta$.[36] Experimental measurement of flexoelectric effects and their theoretical prediction are challenging, and the flexoelectric response of dielectrics, beyond the classical proper ferroelectrics having the perovskite structure, are limited[28,37-41]. Morevoer, the microscopic picture of flexoelectric coupling between strain gradient and non-polar order parameter beyond Kogan's mechanism is still absent.

Here, we focus on the flexoelectric response of improper ferroelectrics with the hexagonal rare-earth ferrite (h-$R$FeO$_3$, $R$ = Ho-Lu, Y, Sc) structure[42]. The atomic structure of h-$R$FeO$_3$ consists of a triangular lattice of FeO$_5$ bipyramids that are sandwiched between rare earth layers. Below ~1000 K, hexagonal ferrites undergo a coupled non-polar $K_3$ structural distortion and polar $\Gamma_2^-$ distortion from a centrosymmetric $P6_3/mmc$ structure,[43] which features the collective rotation of FeO$_5$ bipyramids and the buckling of the rare earth layers. The coupling between the $K_3$ distortion and the polar $\Gamma_2^-$ mode with imbalanced displacements of rare-earth layer along the $c$ axis leads to a spontaneous polarization ($\approx$ 10 $\mu$C/cm$^2$) and the improper ferroelectric order (see **Fig. 1a**)[44–46]. Below ~150 K, assisted by the $K_3$ distortion mode, hexagonal ferrites develop a 120-degree antiferromagnetic order with weak ferromagnetism, making them multiferroic.[47–49] Due to the improper nature of ferroelectricity, h-$R$FeO$_3$ have small piezoelectric coefficients[50], and are less susceptible to depolarization field, which makes them advantageous to achieve large flexoelectric effects at the nanoscale, unlike proper ferroelectrics where the polarization is quenched at nanoscale dimensions[44]. If a gaint strain gradient can be realized in h-$R$FeO$_3$, intrinsic flexoelectricity can be extracted from the variation of the $K_3$ structural distortion, which consists of large atomic displacements ($\approx$ 0.5 Å)[44,46], given that the relationship between $K_3$ distortion and polarization has been well established.

In this work, we demonstrate a novel mechanism of flexoelectricity based on improper ferroelectricity in h-YbFeO$_3$, which we call improper flexoelectricity. Using atomic resolution scanning transmission electron miscope (STEM) imaging, we show that near both grain boundaries and edge dislocation, the large strain gradient (~ $10^6$ m$^{-1}$) strongly suppresses the magnitude of the $K_3$ distortion, which is consistent with the phenomenological Landau theory-based model of improper ferroelectricity. The observed correlation between strain gradient and phases of the $K_3$ distortion or polarization directions originates from the stable and metastable state on energy landscape, suggesting a bias imposed by the strain gradient which breaks the inversion symmetry. The flexoelectric coefficient (~10 nC/m) that is 1-2 orders of magnitude larger than that estimated from the Kogan's mechanism for h-YbFeO$_3$.

The $K_3$ structural distortion in h-$R$FeO$_3$ can be described with a magnitude $Q$, and the direction of in-plane displacement of the apical oxygen atoms using an angle (or phase) $\phi$, as shown in **Fig. 1b**. The free energy of the hexagonal ferrites with the double order parameters, $Q$ and $\phi$, can be written as:[51]

$$F = \frac{a}{2}Q^2 + \frac{b}{4}Q^4 - gQ^3 P \cos(3\phi) + \frac{g_P}{2}Q^2 P^2 + \frac{a_P}{2}P^2, \qquad (1)$$

where $Q$ and $\phi$ are the magnitude and the phase of the $K_3$ distortion, $P$ is the magnitude of the polar distortion ($\Gamma_2^-$) that is proportional to polarization $\mathcal{P}$, and $a$, $b$, $g$, and $g'$ are coefficients. As the first two terms dominate, minimization of the free energy with respect to $Q$ leads to $Q_{min} \approx Q_0 = \sqrt{\frac{-a}{b}}$, where $a < 0$, $b > 0$. The next two terms in Eq. (1) describe the coupling between the $\Gamma_2^-$ and the $K_3$ structural distortions. Minimization of the free energy with respect to $P'$ leads to:

$$P_{min} = \frac{gQ_{min}^3 \cos(3\phi)}{a_P + g_P Q_{min}^2}, \qquad (2)$$

The calculated energy landscape (using the parameters in Ref.[50]) has six energy minima, corresponding to $\phi = 2n\pi/3$ ($n$ is integer) for $P_{min} = P_0$ and $\phi = (2n+1)\pi/3$ for $P_{min} = -P_0$, as shown in **Fig. 1c**, where $P_0 = \frac{gQ_0^3}{a_P + g_P Q_0^2}$.

Depending on the nature of the strain gradient, it can break different inversion symmetries and change the magnitude $Q$ of the $K_3$ structural distortion in h-YbFeO$_3$. Given the strong coupling of $K_3$ mode with the polar $\Gamma_2^-$ mode, we look into the effect of a strain gradient on the energy landscape described by the Landau model. The corresponding free energy can be written as (see details in Supplementary Materials Section 2):

$$F_{flexo} = \frac{f_1}{2} Q^2 \eta^2 + f_2 Q^3 \cos(3\phi)\eta, \qquad (3)$$

where $\eta$ is a general symbol for strain gradient, $f_1$ and $f_2$ are coupling coefficients. The addition of the term $\frac{f_1}{2}Q^2\eta^2$ to Eq. (1) changes the $K_3$ distortion magnitude to:

$$Q_{min} = \sqrt{\frac{-a - f_1 \eta^2}{b}}, \qquad (4)$$

for the minimum free energy. The calculated energy landscape in **Fig. 1d** shows that, with $f_1 > 0$, the energy minima occur at $Q_{min} < Q_0$. The term $f_2 Q^3 \cos(3\phi)\eta$ in Eq. (3) breaks the symmetry between $P_{min} > 0$ and $P_{min} < 0$ in the energy landscape. With $f_1 = 0$ and $f_2 > 0$, by combining Eq. (1) and Eq. (3), one finds

$$Q_{min} = \frac{-3f_2\eta\cos(3\phi)+\sqrt{[3f_2\eta\cos(3\phi)]^2-4ab}}{2b}. \tag{5}$$

In other words, for strain gradient $\eta$, $Q_{min}$ also depends on the phase $\phi$. The calculated energy landscape in **Fig. 1e** shows that, with $f_2\eta < 0$, the three energy minima $\phi = (2n+1)\pi/3$ ($P_{min} < 0$) become local minimum with $Q_{min} < Q_0$ while the other three energy minima $\phi = 2n\pi/3$ ($P_{min} > 0$) remain global with $Q_{min} > Q_0$. This creates a bias of positive $P_{min}$ over negative $P_{min}$, as if the strain gradient ($\eta > 0$) was an electric field. Similarly with $f_2\eta > 0$, negative $P_{min}$ is favored.

Assuming $f_1 = 3.7\times10^7$ eV/f.u. and $f_2 = -1\times10^3$ eV/(Å² f.u.) based on the atomic-resolution measurements as shown below and Landau parameters in Ref [51], the dependence of $Q_{min}$ on $\eta$ is calculated and plotted in **Fig. 1f**. The strain gradient $\eta$ splits the $Q_{min}$ values into a higher stable branch (in solid line) and a lower metastable branch (in dash line). $Q_{min}$ becomes zero at $\eta_0 = \sqrt{-\frac{a}{f_1}}$, corresponding to total quenching of the K₃ distortion. **Fig. 1g** is the calculated $\eta$ dependence of $P_{min}$. Again, the presence of positive $\eta$ reduces the negative $P_{min}$ branch more effectively than it does on the positive $P_{min}$ branch. For $\eta f_2 < 0$, the positive $P_{min}$ branch is favored. Therefore, in improper ferroelectric hexagonal ferrites or isormorphic manganites, it is expected that by coupling to structural distortions, strain gradient $\eta$ can reduce the structural distortion and favor a certain direction of polarization, which leads to an improper flexoelectric effect.

Large strain gradients can only exist in regions where the crystal undergoes dramatic spatial changes such as near dislocations and grain boundaries [19,21,22,52]. As illustrated in **Fig. 2a**, near a grain boundary, the basal plane of the h-$R$FeO₃ unit cells is bent. The deviation of the $c$ axis (polar axis) from the vertical direction can be described by a small bending angle $\theta$, which is approximately equal to the shear strain $e_{zx}$. $\theta$ is larger near the grain boundary and smaller away from the boundary, suggesting a strain gradient $\eta = \frac{\partial e_{zx}}{\partial x}$. The schematic atomic structure for the bent half-layer is given in **Fig. 2b**. **Fig. 2c** shows a high-angle annular dark-field scanning transmission electron microscope (HAADF-STEM) image of the atomic structure of h-YbFeO₃ near a grain boundary viewed along the [100] direction. The existence of YbFe₂O₄ stacking faults, which are highlighted by dashed rectangular boxes, as well as the long-range bending of h-YbFeO₃ can be identified near the grain boundary due to vertical mismatch. Close-up views of a strongly and a weakly bent region are given in I and II, respectively. The comparison between area I and area II confirms the variation of $\theta$: larger $\theta$ near the grain boundary and smaller $\theta$ far away from the boundary. It should be noticed that though the multidomain or topological strip domain states are formed under the influence of shear strain[51,53] (see spatial mapping of $\phi$ in **Fig.S10 b**), the polarization is favored to up direction in strong bending region near grain boundary.

Using the FeO layer, which is flat in the unstrained structure (unlike the buckled Yb layer), we calculated the $\theta$ values as a function of position (see Supplementary Materials Section 3). It turns out that $\theta$ depends primarily on the position $x$ (along [-211] direction). In particular, the difference in $\theta$ in the regions I and II is $4 \pm 0.5$ deg (see **Fig.3 a**). Furthermore, we calculated local strain gradient $\eta = \frac{\partial e_{zx}}{\partial x} = \frac{\partial \theta}{\partial x}$. The results are shown in **Fig. 3b** as a 2D map, where the maximum value of $\eta$ reaches $6\times10^6$ m$^{-1}$.

The corrugation of the rare earth layers ($Q'$) is a representative component of the $K_3$ distortion ($Q' \approx 0.4\ Q$)[54,55]. In **Fig. 3c**, $Q'$ values are calculated and overlaid on the image for the area with bent lattice. Clearly, $Q'$ values are smaller in the bent lattice compared to the regions that don't have a noticeable strain gradient (see **Fig.S10 a**). More specifically, from right to left, as the strain gradient increases (see **Fig. 3b**), $Q'$ decreases from $\approx 35$ pm to less than 10 pm, suggesting that the strong strain gradient is correlated with the suppression of $Q'$. **Fig. 3d** displays the $Q'$ and ϕ values of flat lattice in a polar plot. Here ϕ values are concentrated around 120⁰, indicating a single domain of positive polarization (see **Fig. 1c**). The $Q'$ values have a narrow distribution around 40 pm. On the contrary, in the polar plot **Fig. 3e** for the bent region, a clear distribution of both $Q'$ and ϕ is visible. In addition to bending regions near grain boundries, the similar magnitude of bending angle and strain gradients also exist near edge dislocation (see Supplementary Materials Section 3, 4), associated suppression of $Q'$ indicates the effect of improper flexoeletricity is independent of the type of defects which induces the bending.

The quantitative correlation between $Q'$ and $\eta$ is plotted in **Fig. 4a**; the bent areas come from two different regions near the grain boundary and edge dislocation (see **Fig. S2** and **Fig. S6**). The suppression of $Q'$ in the severely bent region indicates a significant effect of the $f_1$ term in $F_{\text{flexo}}$. By fitting the observation in **Fig. 4a** using Eq. (4) and $Q' \approx 0.4 Q_{\min}$, one finds $f_1 = (3.7 \pm 0.5) \times 10^7$ eV/f.u. The quantitative correlation between $P$ and $\eta$ can be calculated using Eq. (2). The results are plotted in **Fig. 4b**. Overall, $P$ decreases with $|\eta|$, which is due to the reduction of $Q'$. Electric polarization can be roughly estimated using $P = P_0 \frac{P}{P_0}$. Here we assume $P_0 = 10$ μC/cm$^2$ found experimentally for the unstrained lattice of h-YbFeO$_3$ [42,53], $P_0$=0.16 Å according to first-principles calculations [56]. With the polarization estimated, one can calculate average flexoelectric coefficient $\bar{\mu} = \frac{\Delta P}{\Delta \eta}$. According to **Fig. 4b**, polarization is reduced to zero with a strain gradient $\eta_0 = 2.5 \pm 0.5 \times 10^6$ m$^{-1}$. Hence, one finds average $\bar{\mu} \approx \frac{P_0}{\eta_0} = 40 \pm 10$ nC/m.

Notice that the flexoelectric effect reported here is an indirect effect due to the coupling between strain gradient $\eta$ and the magnitude $Q$ of the non-polar $K_3$ structural distortion, described with the $f_1$ term in Eq. (3). The clear $\eta$ dependence allows for the determination of $f_1$. The effect of the $f_2$ term is, however, mainly to split $Q$ and $P$ into a stable branch and a metastable branch. As shown in **Fig. 4b**, when the magntitued of strain gradient is larger than $\sim 2 \times 10^6$ m$^{-1}$, the large positive (negative) strain gradient tends to favor the positive (negative) $P$, which is consistent with the theoretical expectation for the biasing effect of strain gradient which breaks the inversion symmetry. Moreover, under lower strain gradient, the distribution of polarization states on metastable branch can be interpreted as the convolution of improper flexoelectricity and shear strain-induced topological stripe domain state[51,53], where the suppression of $P$ still follows the model.

One can estimate the flexoelectric coefficient using Kogan's method,[35] which says $\mu = f'\chi$, where $\chi$ is the electric susceptibility and $f'$ is the electric potential generated per unit strain gradient. Kogan derived that $f'$= 1-10 V for most materials.[35] Taking $\chi \approx 30\ \varepsilon_0$ for h-YbFeO$_3$[44], where $\varepsilon_0$ is the vacuum permittivity, one finds that μ = 0.3-3 nC/m. This result is one or two orders of magnitude smaller than the $\bar{\mu}$ value estimated above using experimentally observed atomic displacements, confirming that the improper flexoelectricity is distinct from the generic Kogan mechanism.

In summary, the nanoscopic large strain gradient was observed in h-YbFeO$_3$ films near the grain boundaries and edge dislocation, and the correlation between the strain gradient and the primary order parameters suggests a giant flexoelectric effect in h-YbFeO$_3$. Based on a phenomenological model, at least two coupling mechanisms are identified, in which strain gradient suppresses non-polar K$_3$ structiral distortion and biases the secondary polarization, respectively. The improper flexoelectricity, corresponding to the indirect electromechanical coupling between polarization and strain gradient, offers new insight into the microscopic origin of flexoelectric effects in improper ferroelectric hexagonal ferrites or manganites beyond conventional Kogan's mechanism.

**Acknowledgements**

Funding: This work was primarily supported by the National Science Foundation (NSF), Division of Materials Research (DMR) under Grant Nos. DMR-1454618 and DMR-2145797, and by the Nebraska Center for Energy Sciences Research. The research was performed in part in the Nebraska Nanoscale Facility: National Nanotechnology Coordinated Infrastructure and the Nebraska Center for Materials and Nanoscience, which are supported by the NSF under Grant No. ECCS- 2025298, and the Nebraska Research Initiative. The Microscopy work was conducted as part of a user project at the Center for Nanophase Materials Sciences (CNMS), which is a US Department of Energy, Office of Science User Facility at Oak Ridge National Laboratory.

**Author contributions**

The thin film synthesis and x-ray diffractions were carried out by X.L. and Y.Y. under the supervision of X.X. (S)TEM experiments were conducted by G.R. and A.S.T. under the supervision of R.M. The study was conceived by X.L. and X.X. X.L., G.R., Y.Y, R.M,and X.X. co-wrote the manuscript. All the authors discussed the results and commented on the manuscript.

**Competing interests**

The authors declare no competing interest.

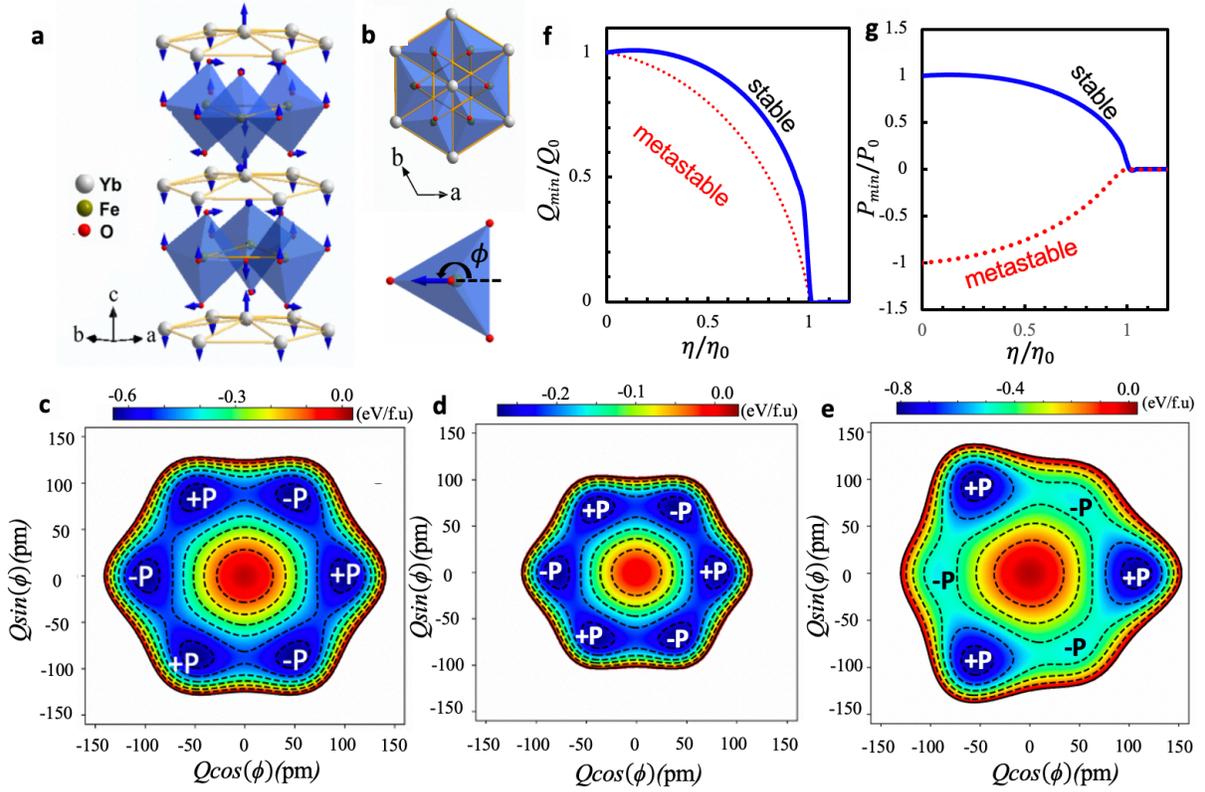

**Fig. 1** Phenomenological model of flexoelectricity in h-RFeO3. (a) Atomic structure of the improper ferroelectric h-YbFeO3. The arrows indicate the displacement of atoms. (b) In-plane view of h-$R$FeO3 and the displacement of apical oxygen within FeO5 bipyramid, corresponding to order parameters $(Q, \phi)$. The energy landscape with $f_1 = f_2 = 0$ in (c), $f_1 = 3.7 \times 10^7$ eV/f.u., $f_2 = 0$, and $\eta = 1 \times 10^6$ m$^{-1}$ in (d), and $f_1 = 0, f_2 = -1 \times 10^3$ eV/(Å$^2$ f.u.), and $\eta = 1 \times 10^6$ m$^{-1}$ in (e) respectively. (f) and (g) are the dependence of $Q_{min}$ and $P_{min}$ (see text) on the strain gradient $\eta$, assuming $f_1 = 3.7 \times 10^7$ eV/f.u. and $f_2 = -1 \times 10^3$ eV/(Å$^2$ f.u.).

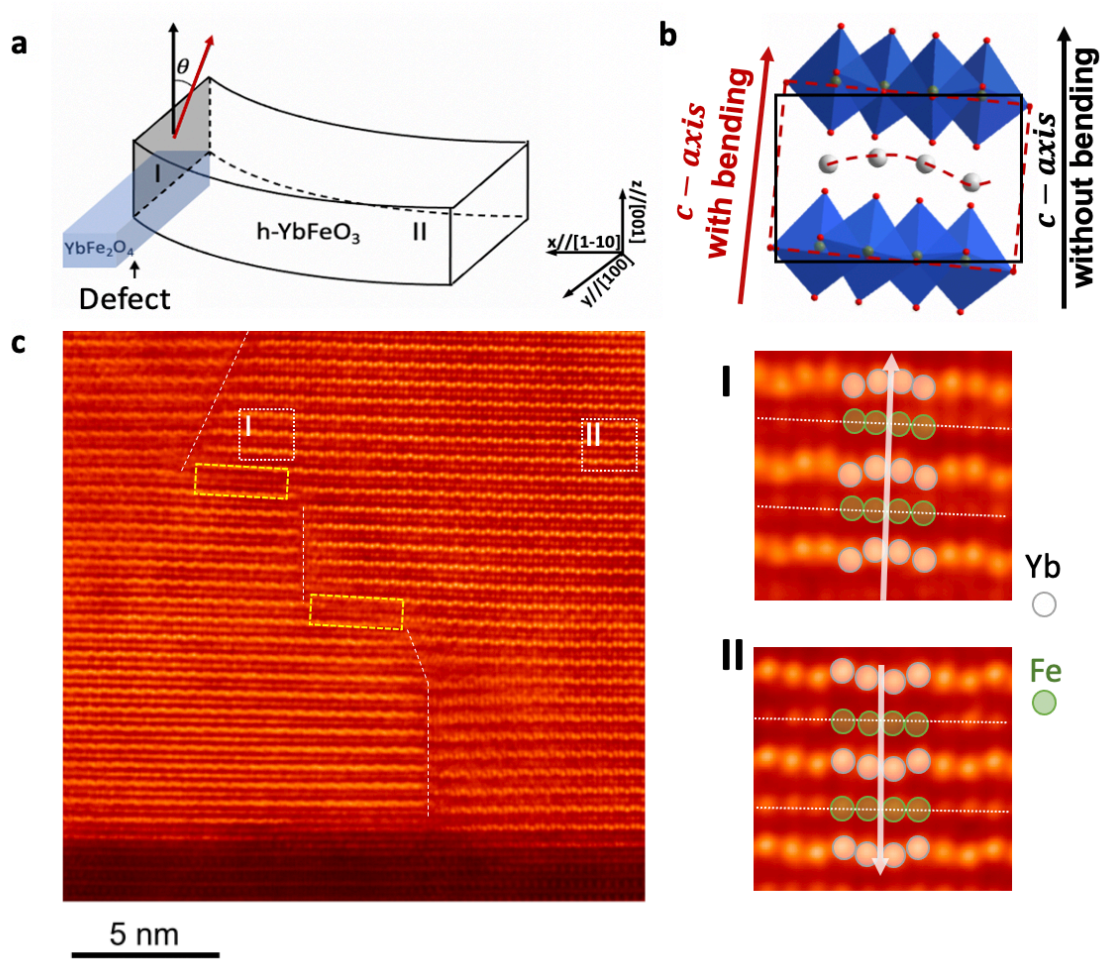

**Fig. 2** (a) Schematic of large bending of h-YbFeO$_3$ due to a h-YbFe$_2$O$_4$ grain. $\theta$ is the bending angle. (b) The schematic atomic structure of half-layer of h-YbFeO$_3$ without and with bending. (c) HAADF-STEM image of h-YbFeO$_3$ thin films showing of an area around a grain boundary with YbFe$_2$O$_4$ defects. The white dashed line corresponds to the grain boundary and yellow dashed rectangular boxes highlight YbFe$_2$O$_4$ defects. Area I and II are close-up views of the areas in (c) indicated by the white dashed boxes. The arrows indicate the polarization direction.

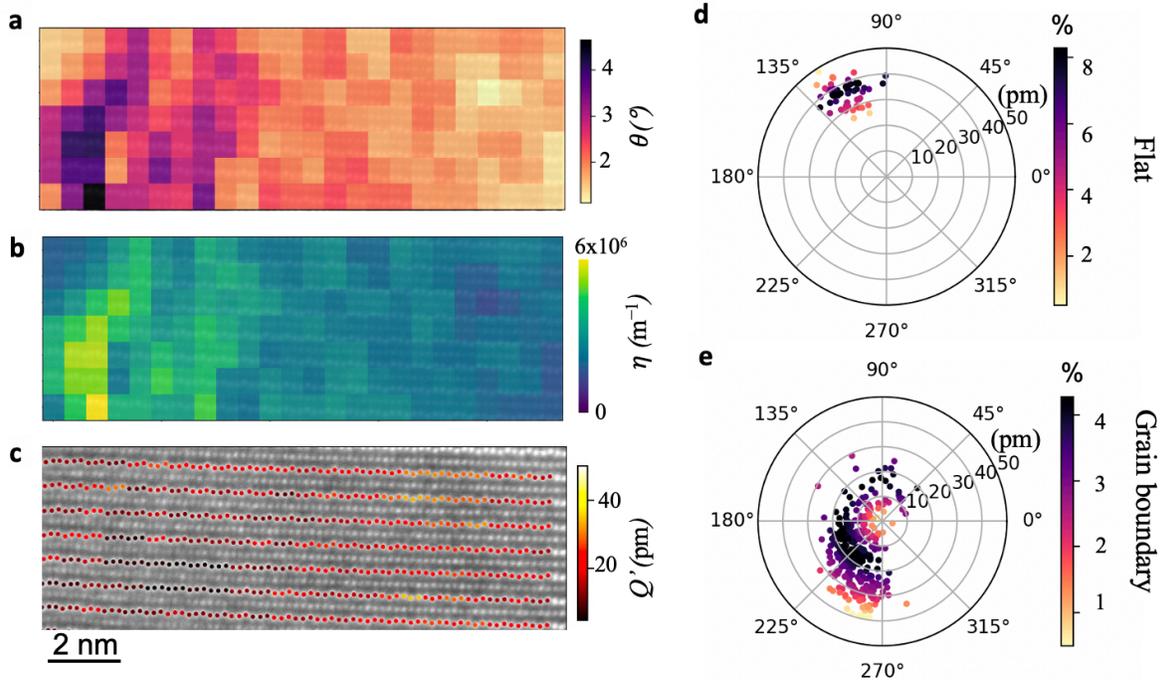

**Fig. 3** (a) Spatial mapping of bending angle, (b) strain gradient, and (c) order parameter $Q'$ in the bending region near a grain boundary in h-YbFeO$_3$. (d) and (e) are pole figures of ($Q'$, $\Phi$) for the flat region and and bending region of (c), respectively.

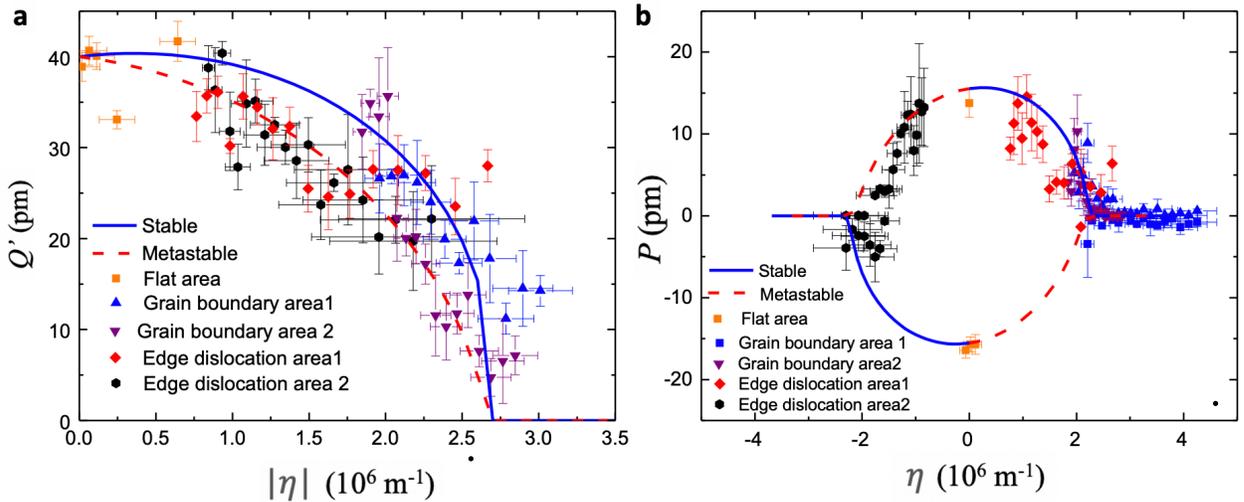

**Fig. 4** The dependence of $Q'$ (a) and $P$ (b) on the strain gradient and the fitting with the improper flexoelectricity model.